\documentclass{emulateapj}

\shorttitle{The S2-orbit from VLT and Keck data}
\shortauthors{Gillessen et al.}

\begin{document}

\title{The orbit of the star S2 around Sgr~A* from VLT and Keck data}

\author{S.~Gillessen\altaffilmark{1}, F.~Eisenhauer\altaffilmark{1}, T. K.~Fritz\altaffilmark{1}, H.~Bartko\altaffilmark{1}, K.~Dodds-Eden\altaffilmark{1}, O.~Pfuhl\altaffilmark{1}, T.~Ott\altaffilmark{1}, R.~Genzel\altaffilmark{1,2}}
\altaffiltext{1}{Max-Planck-Institut f\"ur Extraterrestrische Physik, 85748 Garching, Germany}
\altaffiltext{2}{Physics Department, University of California, Berkeley, CA 94720, USA}

\begin{abstract}
Two recent papers (Ghez et al. 2008, Gillessen et al. 2009) have estimated the mass of and the distance to the massive black hole in the center of the Milky Way using stellar orbits. The two astrometric data sets are independent and yielded consistent results, even though the measured positions do not match when simply overplotting the two sets. In this letter we show that the two sets can be brought to excellent agreement with each other when allowing for a small offset in the definition of the reference frame of the two data sets. The required offsets in the coordinates and velocities of the origin of the reference frames are consistent with the uncertainties given in Ghez et al. (2008). The so combined data set allows for a moderate improvement of the statistical errors of mass of and distance to Sgr~A*, but the overall accuracies of these numbers are dominated by systematic errors and the long-term calibration of the reference frame. We obtain $R_0 = 8.28 \pm 0.15 |_\mathrm{stat} \pm 0.29|_\mathrm{sys}\,$kpc and $M_\mathrm{MBH} = 4.30 \pm 0.20 |_\mathrm{stat} \pm 0.30|_\mathrm{sys} \times 10^6\,M_\odot$
as best estimates from a multi-star fit.
\end{abstract}


\section{Introduction}
The motions of stars in the immediate vicinity of Sgr~A* have been tracked since 1992 at the NTT/VLT and since 1995 at the Keck telescope \citep{Eckart:1996p163,Ghez:1998p118}. With the detection of accelerations \citep{Ghez:2000p138} and the determination of the first orbit \citep{Schodel:2002p153} these measurements provided firm evidence for the existence of a massive black hole (MBH) at the center of the Milky Way, coincident with the radio-source Sgr~A*. The stars are used as test particles for the gravitational potential in which they move. Particularly important is the bright ($m_\mathrm{K}\approx14$) star S2 (S0-2 in the Keck nomenclature) orbiting the MBH in 15.9 years on an ellipse with an apparent diameter of $\,\approx0.2''$. Together with radial velocity data, the astrometric data allow for a geometric determination of $R_0$, the distance to the Galactic Center \citep{Salim:1999p1362,Eisenhauer:2003p128}. The number of orbits has increased to more than 20 since then \citep{Ghez:2005p1681,Eisenhauer:2005p117,Gillessen:2009p1117} and in particular the full S2 orbit has recently been used for improved determinations of the mass and of $R_0$ \citep{Ghez:2008p945,Gillessen:2009p1117}. 

These two astrometric data sets are independent since they were obtained at different telescopes and were analyzed by different teams using different tools. On the other hand, the radial velocity data attached to the astrometric data sets largely overlap, mainly because this technically is straight-forward. The radial velocities refer to the local standard of rest and hence the inclusion of other data into a given data set needs no special care. The situation is different for astrometric data, for which only an approximate realization of an absolute reference frame currently is possible \citep{Reid:2007p169}. This means that the exact definition of coordinates is a matter of the respective data analysis, and simply merging two lists of astrometric positions will fail (see figure~\ref{f1}).

The two groups \citep{Ghez:2008p945,Gillessen:2009p1117} come to very similar conclusions for the mass of Sgr~A* and $R_0$. From VLT data \cite{Gillessen:2009p1117} derived (when using the S2 orbit only)
\begin{eqnarray}
R_0&=&8.31 \pm 0.33|_\mathrm{stat} \,\mathrm{kpc}\nonumber\\
M_\mathrm{MBH} &=& 4.29 \pm 0.07|_\mathrm{stat} \pm 0.34|_{R_0}\,M_\odot\,\,
\end{eqnarray}
where the statistical error on the mass is for a fixed distance and the second error term is due to the statistical error on $R_0$. The latter includes already the coordinate system uncertainty. In addition, a systematic error of $\Delta R_0 \approx \pm^{\,0.5}_{\,1.0}\,$kpc is present, owing to the uncertainties of a) how much the 2002 data (close to periastron) can be trusted and b) the assumption that the effective potential is Keplerian. The systematic error of $R_0$ also influences $M_\mathrm{MBH}$ since the two quantities are correlated.
\cite{Ghez:2008p945} obtained from the Keck data set (neglecting 2002 data, and fixing the radial velocity $v_z$ of the MBH to 0)
\begin{eqnarray}
R_0&=&8.4 \pm 0.4\,\mathrm{kpc} \nonumber\\
M_\mathrm{MBH} &=& 4.5 \pm 0.4\,M_\odot\,\,.
\end{eqnarray}
The errors are of statistical nature and the mass error includes the uncertainty induced by the error on $R_0$. Probably, also a systematic error should be taken into account here. Leaving $v_z$ free yielded somewhat larger statistical errors, e.g. $\Delta R_0=0.6\,\mathrm{kpc}$.

The aim of this letter is to investigate the combined data set. We show that not only the fit results are in excellent agreement, but also that the combined data can be described very well by a single Keplerian orbit. The such combined data set also yields an improvement in the statistical fit errors
for the derived parameters. This can be understood qualitatively by inspecting the data: Before 2002, the Keck measurements appear superior to our NTT/VLT data set. At that time, our group was using the NTT with an aperture of $3.6\,$m, which is notably smaller than $10\,$m for the Keck telescopes. After 2002, the VLT data has comparable errors to the Keck data for individual data points but a much denser time sampling. This is owed to the location on the southern hemisphere of the VLT and the generous allocation of telescope time to the Galactic Center projects. Hence, by suitably combining the two data sets, one should obtain a data set which combines the respective advantages.\footnote{We use data collected between 1992 and 2009 at the European Southern Observatory, both on Paranal and La Silla, Chile; in particular from the Large Programs 179.B-0261 and 183.B-0100.}

\section{Data Base}

\begin{figure}[htbp]
\begin{center}
%\plotone{f1.pdf}
\plotone{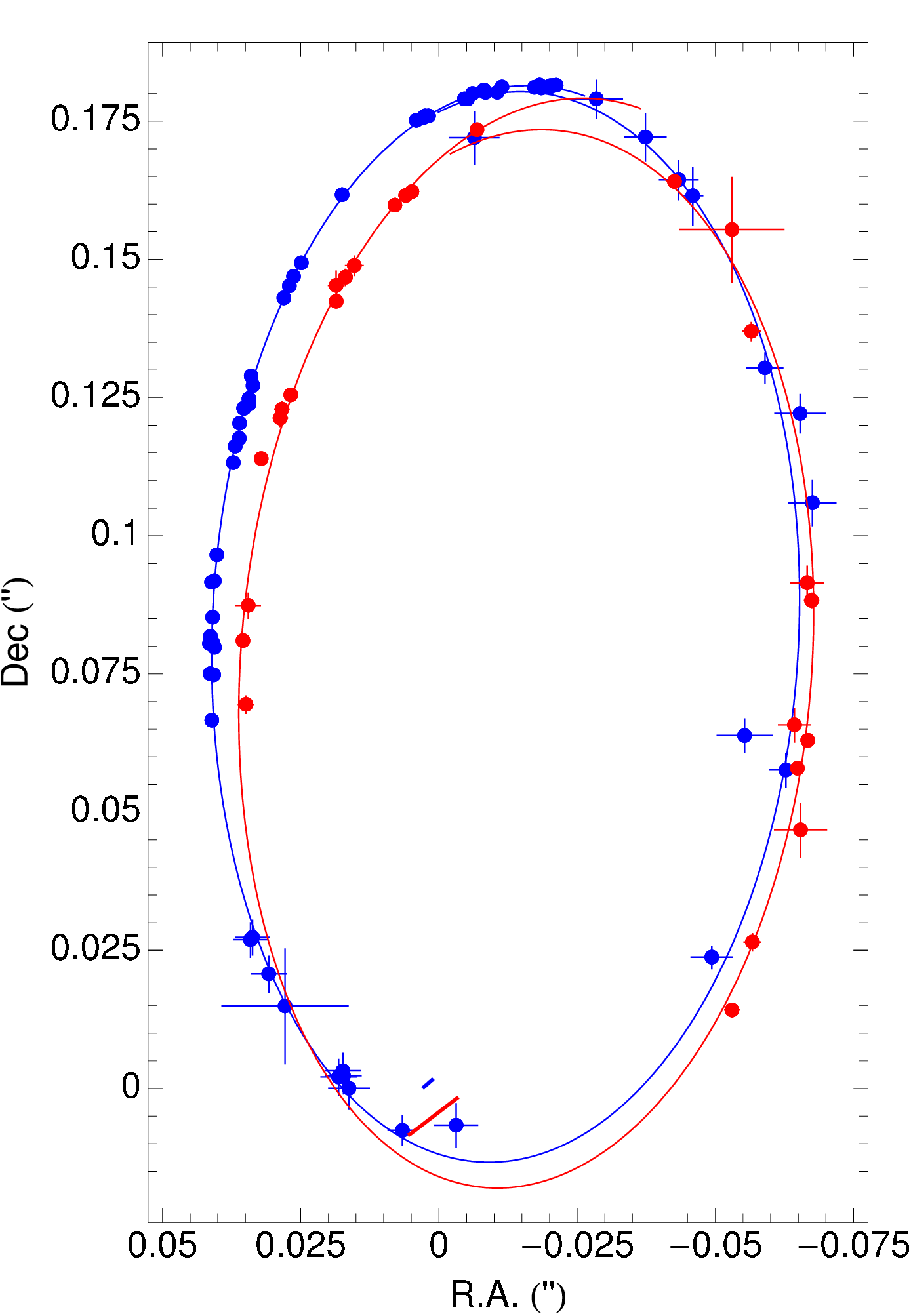}
\caption{Astrometric data for the star S2. Blue: NTT/VLT measurements. Red: Keck measurements. The lines show the Keplerian orbit fits for the respective data set, which do not yield closed ellipses in this figure due to the motion of the point mass with respect to the chosen coordinate systems. The small lines close to the origin indicate the position of the center of mass as a function of time.}
\label{f1}
\end{center}
\end{figure}

\begin{figure}[htbp]
\begin{center}
%\plotone{f2.pdf}
\plotone{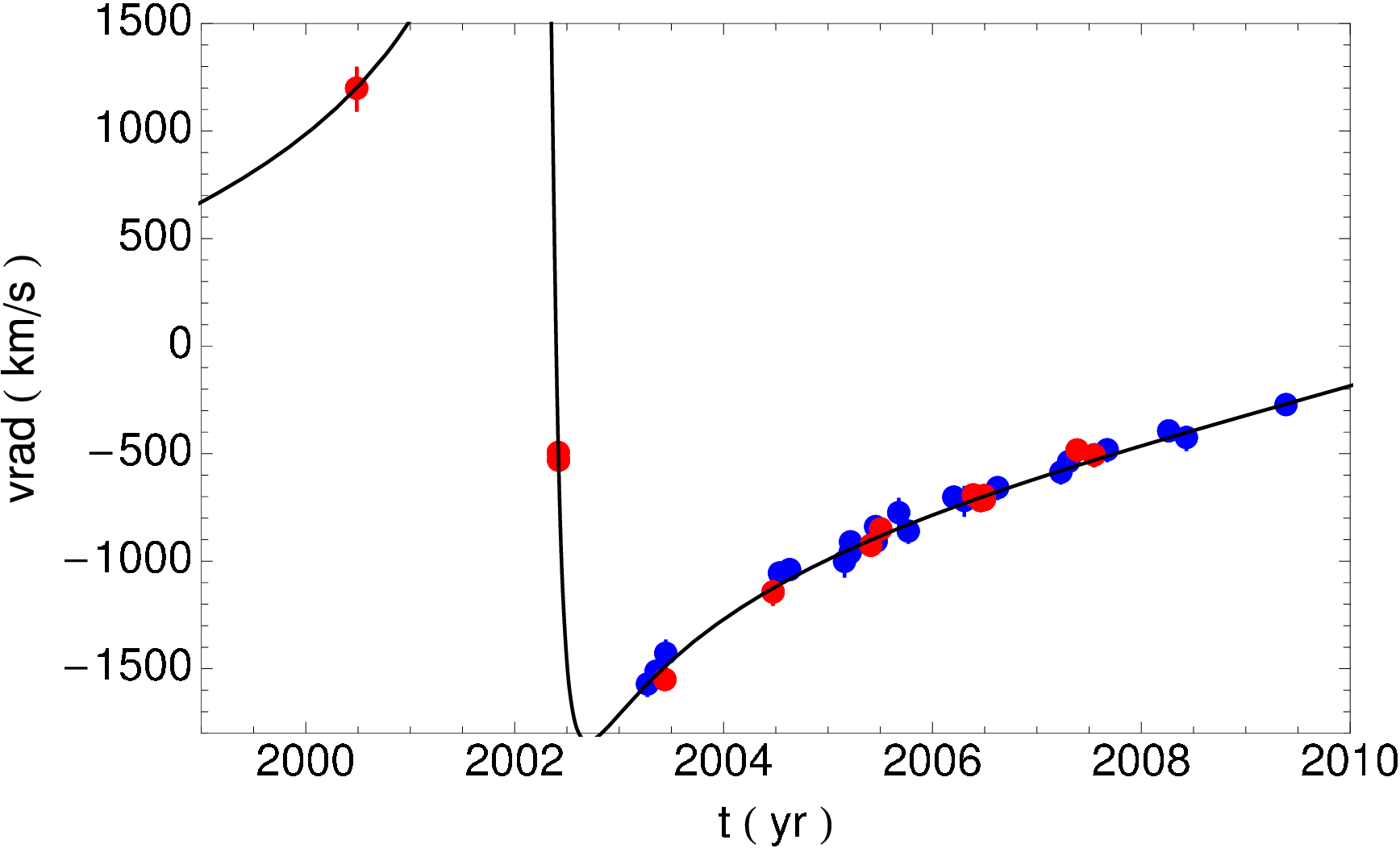}
\caption{Radial Velocity data for the star S2. Red: Keck measurements. Blue: VLT measurements. The black line shows already the result of a combined orbit fit.}
\label{f2}
\end{center}
\end{figure}

The data used here is shown in figures~\ref{f1} and~\ref{f2}. We resign a detailed discussion of the data, since it is well presented in the original works and would be beyond the scope of this letter. 

The astrometric data from the NTT/VLT we are using are identical to the data from \cite{Gillessen:2009p1117} up to August 2008. Since then, one more epoch has been observed in 2008 and 11 more epochs in 2009. 
We treated these data in precisely the same way as described in \cite{Gillessen:2009p1117}, where the somewhat cumbersome derivation of astrometric positions from the imaging data is described in detail. As done previously, we assigned lower weights to the VLT data from 2002 to account for the fact that the star might have been confused in that year.
In total, the NTT/VLT data set contains 70 epochs from March 1992 to August 2009. Furthermore, we added one more epoch (March 2009) to the measurements of reference stars by which the coordinate system is tied to the ICRF. 

\cite{Ghez:2008p945} presented in their table 3 the measured positions for all Keck epochs\footnote{The definition of their $X$-coordinate (increasing values to the east), however, does not match the signs of the numbers in the table, which have to be reverted (S2 was West of Sgr~A* in the 1990s and was East of Sgr~A* from 2003 to 2007).}. We use the given 26 epochs, ranging between 1995 and 2008. 

For the radial velocities, we use the combined data base from \cite{Gillessen:2009p1117} and \cite{Ghez:2008p945}\footnote{In table 4 of the latter paper two radial velocity measurements  from the year 2002 are missing. They are shown in the figures and clearly have been used in the fit. The values can be found in \cite{Ghez:2003p178} though.}. Finally, we add one more VLT epoch from 2009, which brings the total number of radial velocity points to 36.

\section{Orbital Fits}
Figure~\ref{f1} shows that the two astrometric data sets cannot be put together in a simple fashion. 
To account for the difference we make the following, simple assumption:
The two data sets only differ in their definition of the coordinate system, namely the position of the origin and the zero velocity\footnote{A difference in rotation is very unlikely, since the images can be oriented very well by the means of stars across the field of view.}. This yields four parameters $(\Delta x,\,\Delta y,\, \Delta v_x,\, \Delta v_y)$ by which the difference can be described. Given that both data sets contain many more data points, it is feasible to solve for these four parameters during the fitting procedure. These four parameters come hence in addition to the six orbital elements and the parameters describing the gravitational potential ($M_\mathrm{MBH}$, $R_0$, position, zero velocity, $v_z$). 

\begin{figure}[htbp]
\begin{center}
%\plotone{f3.pdf}
\plotone{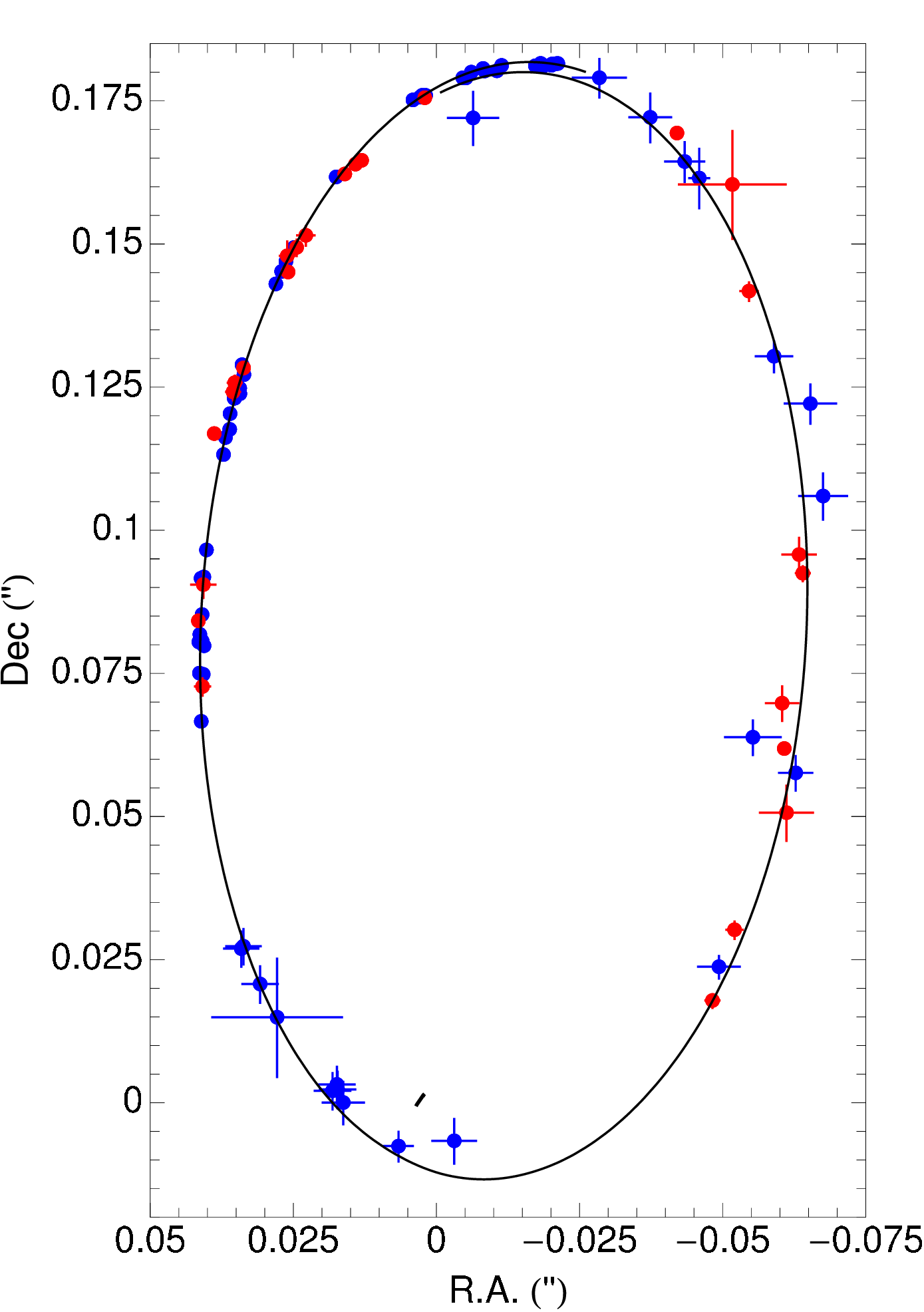}
\caption{Result of the combined orbit for the star S2. Blue: NTT/VLT measurements. Red: Keck measurements. The black line shows the Keplerian fit (row 1 in table~\ref{fitres}). }
\label{f3}
\end{center}
\end{figure}

We have implemented the additional four parameters in the code of \cite{Gillessen:2009p1117}. For (most of) the orbital fits, we used the same assumptions as for the preferred fit in \cite{Gillessen:2009p1117}, namely we imposed priors on the NTT/VLT coordinate system definition which reflect our best estimate of how well an absolute coordinate system was established: 
\begin{equation}
\begin{array}{cc}
\mathcal{P}(x) = 0 \pm 1.0 \,\mathrm{mas}  \,& \,\,\,\,\,\,\mathcal{P} (v_x) =0 \pm 0.1 \,\mathrm{mas/yr} \nonumber \\
\mathcal{P} (y) =0 \pm 2.5\,\mathrm{mas} \,&\,\,\,\,\,\, \mathcal{P} (v_y) =0 \pm 0.1 \,\mathrm{mas/yr}  \nonumber \\
&\mathcal{P} (v_z) =0 \pm 5\,\mathrm{km/s}\,\, .
\end{array}
\label{priorsUsed}
\end{equation}
Using that, the combined data set can be described very well by a Keplerian orbit (figure~\ref{f3}). 
For the 211 degrees of freedom (96 astrometric epochs which count twice and 36 radial velocity points minus 17 parameters) we obtained $\chi^2 = 271$ for the fit shown in figure~\ref{f3}, corresponding to a reduced $\chi^2$ of 1.28.

The fit result for the four additional parameters is
\begin{eqnarray}
\Delta x &=& -3.7 \pm 0.6 \,\mathrm{mas}  \nonumber \\
\Delta y &=& -4.1 \pm 0.6\,\mathrm{mas} \nonumber \\
\Delta v_x &=&-0.68 \pm 0.11 \,\mathrm{mas/yr} \nonumber\\
\Delta v_y &=&0.26 \pm 0.10 \,\mathrm{mas/yr} \,\,.
\label{csysdiff}
\end{eqnarray}
We conclude that these parameters are well-determined from the fit, which validates our approach for combining the two data sets. The positional difference $(\Delta x,\, \Delta y)$ is somewhat larger than the actual mismatch between the two coordinate systems. This happens simply since $(\Delta x,\, \Delta y)$ refer to the epoch May 2005, the zero point in time of the NTT/VLT coordinate system, while the Keck data uses July 2004 for that. Nevertheless it is worth checking how well the numbers in equation~\ref{csysdiff} compare to the expected accuracies of the coordinate systems. The uncertainty of the NTT/VLT coordinate system (equation~\ref{priorsUsed}) is smaller than the values in equation~\ref{csysdiff}. \cite{Ghez:2003p178} conclude that their coordinate system is accurate to $(\Delta x,\, \Delta y) = \pm (5.7,\,5.7)\,$mas in position and
to $(\Delta v_x,\, \Delta v_y) = \pm (0.6,\,0.9)\,$mas/yr. This is consistent with equation~\ref{csysdiff}. The larger uncertainty of the Keck coordinate system is due to the shorter timeline used for determining the motions of the reference stars ($\approx 2$ years compared to $\approx6$ years for the NTT/VLT data).
\begin{table}
\caption{\label{s2params} Best fitting orbital elements for S2 from a combined orbit fit. The errors are rescaled such that the reduced $\chi^2$ is 1. The corresponding gravitational potential is given in the first row of table~\ref{fitres}.}
{\scriptsize 
\begin{center}
\begin{tabular}{lc}
Parameter & Value\\
\hline
$a$ & $0.1246'' \pm 0.0019''$\\
$e$ & $0.8831 \pm 0.0034$\\
$i$ &$134.87^\circ \pm 0.78^\circ$\\
$\Omega$ &$226.53^\circ  \pm 0.72^\circ $\\
$\omega$ &$64.98^\circ  \pm 0.81^\circ$ \\
$t_\mathrm{P}$& $2002.3293 \pm 0.0066$
\end{tabular}
\end{center}}
\end{table}

In table~\ref{s2params} we give the orbital elements for the combined fit. The corresponding parameters describing the gravitational potential are presented in the first row of table~\ref{fitres}.
We continued by making different choices for the priors, for the 2002 VLT data and for the model of the gravitational potential. These fits are also listed in table~\ref{fitres}, the respective choices are indicated in each row. 

Comparing the results with table~4 of \cite{Gillessen:2009p1117} one notices that generally the statistical fit errors for $M_\mathrm{MBH}$ and $R_0$ are 20\% to 30\% smaller in the combined orbit fit. Also the errors on the position and 2D velocity of the MBH have decreased moderately. Generally, the fitted position of the MBH has moved by $\approx2.5\,$mas East and $\approx1.5\,$mas North. This is consistent with the accuracy by which the radio source Sgr~A* can be identified in the infrared images.

\begin{table*}
\caption{The central potential from fitting the S2 data. The first six rows use the same assumptions as the first six rows in table~4 of \cite{Gillessen:2009p1117} and thus are directly comparable. The last four rows show post-Newtonian fits. All errors are rescaled such that the reduced $\chi^2$ is 1. The second column is a flag indicating whether the 2002 VLT data from S2 was used. The third column informs which of the priors from equation~\ref{priorsUsed} were used. Column four is a flag whether a relativistic or a Keplerian model was used. The fifth column indicates the radial density profile of the assumed extended mass. The last column is the fitted mass between peri- and apocenter of the S2 orbit divided by $M_\mathrm{MBH}$.  
\label{fitres}}
{\scriptsize
\begin{center}
\begin{tabular}{lcccc|cccccccc}
& VLT & priors &rel.&ext.& $R_0$&$M_\mathrm{MBH} $&
$\alpha $&$\delta $&$v_\alpha $&
$v_\delta $&$v_z$  &$\eta$\\
&2002&&&mass& (kpc) &$(10^6 M_\odot)$ & (mas) & (mas) & ($\mu$as/yr) & ($\mu$as/yr) & (km/s)&\%\\
\hline
\hline
1& yes& 2D, $v_z$ &no&- & $8.34\pm0.27$& $4.40\pm0.27$&$2.93\pm0.55$&$0.41\pm0.85$&$-61\pm84$&$89\pm91$&$2.5\pm6.3$&-\\
2& no & 2D, $v_z$ &no&- & $7.72\pm0.33$& $3.90\pm0.29$&$3.21\pm0.58$&$-1.53\pm1.13$&$-109\pm88$&$67\pm94$&$0.4\pm6.5$&-\\
3& yes & $v_z$ & no&- &$8.39\pm0.28$& $4.44\pm0.29$&$3.18\pm0.62$&$0.14\pm0.88$&$-27\pm115$&$186\pm132$&$2.2\pm6.2$&-\\
4& no & $v_z$ &no&- & $7.67\pm0.37$& $3.87\pm0.31$&$3.72\pm0.66$&$-2.05\pm1.21$&$-134\pm123$&$106\pm138$&$0.24\pm6.4$&-\\
5& yes & none & no&- &$8.77\pm0.41$& $4.99\pm0.53$&$3.13\pm0.62$&$0.41\pm0.88$&$-34\pm115$&$173\pm131$&$33\pm24$&-\\
6& no & none & no&- &$7.73\pm0.57$& $3.93\pm0.58$&$3.69\pm0.67$&$-1.95\pm1.37$&$-131\pm123$&$107\pm138$&$3.7\pm27.0$&-\\
7& yes & 2D &no&- & $8.78\pm0.41$& $5.03\pm0.55$&$2.87\pm0.55$&$0.68\pm0.85$&$-63\pm84$&$81\pm91$&$37\pm24$&-\\
8& no & 2D &no&- & $7.85\pm0.52$& $4.04\pm0.55$&$3.19\pm0.58$&$-1.34\pm1.23$&$-109\pm88$&$66\pm94$&$7.9\pm26.7$&-\\
9& yes & 2D, $v_z$ &yes&- & $8.46\pm0.28$& $4.65\pm0.30$&$2.58\pm0.55$&$-0.10\pm0.84$&$-37\pm84$&$97\pm91$&$1.5\pm6.3$&-\\
10& yes & 2D, $v_z$ &yes&$r^{-1.4}$ & $8.55\pm0.29$& $4.70\pm0.31$&$3.03\pm0.66$&$0.25\pm0.87$&$-130\pm114$&$70\pm94$&$1.4\pm6.3$&$2.0 \pm 1.6$\\
11& yes & 2D, $v_z$ &yes&$r^{-7/4}$ & $8.51\pm0.28$& $4.66\pm0.30$&$3.07\pm0.66$&$0.14\pm0.86$&$-140\pm114$&$67\pm93$&$1.1\pm6.3$&$1.8 \pm 1.4$\\
12& yes & 2D, $v_z$ &yes&$r^{-2.1}$ & $8.49\pm0.28$& $4.64\pm0.30$&$3.09\pm0.66$&$0.00\pm0.85$&$-146\pm115$&$64\pm93$&$0.6\pm6.3$&$1.6 \pm 1.3$\\
\end{tabular}
\end{center}
}
\end{table*}

The comparison also shows that for the combined data set the various fits differ less from each other than for the NTT/VLT-only data set. For example, including or not including the 2002 VLT data (comparing the first and second row) changed $R_0$ in 
\cite{Gillessen:2009p1117} by $0.95\,$kpc, while in the combined data set the same change is only $0.62\,$kpc. Also the sixth row (a fit excluding the 2002 VLT data and neglecting priors) shows that the results are much closer in the combined set. \cite{Gillessen:2009p1117} reported
$R_0=6.63\pm0.91\,$kpc, while the combined data yield $R_0=7.73\pm0.57\,$kpc. The range of values for $R_0$ is in the combined data set $1\,$kpc only, while for \cite{Gillessen:2009p1117} it exceeded $2\,$kpc.

Next, we turned to post-Newtonian orbit models. A relativistic orbit fit is shown in row~9 of table~\ref{fitres}, yielding a slight increase in $R_0$ compared to the Keplerian fit. This bias was already noted by \cite{Zucker:2006p194}. Assuming the relativistic $\beta^2$ effects for the radial velocity \citep{Zucker:2006p194} we then checked if the combined data allow to constrain the relativistic periastron shift. For this purpose we introduced a parameter $f$, which is 0 for a Keplerian model and 1 for the correct relativistic model. The fit yielded a very large error $\Delta f\approx 10$ and we conclude that also the combined data set is not yet sensitive to the relativistic precession induced by the Schwarzschild metric.

We also repeated the tests of \cite{Gillessen:2009p1117} for an extended mass component, using three different power law slopes for the density profile: $\rho(r) \sim r^{-\alpha}$, $\alpha=1.4,\,7/4,\,2.1$. The normalization of the density profile was a parameter, which we determined by the fit. We express the results for this parameter in table~\ref{fitres} by giving $\eta$, the mass between peri- and apocenter of the S2 orbit divided by the fitted mass of the MBH. The results in rows 9-12 are not directly comparable to the numbers in \cite{Gillessen:2009p1117} who imposed an additional prior on $R_0$ (obtained from the other S-stars) for the fits with an extended mass component. The results from the combined data yield nevertheless very similar upper limits: At most very few percent of the central mass can make up an extended mass component inside the S2 orbit.

Overall, the fits presented here show that the chosen scheme to combine the two independent data sets is valid and that consequently (moderately) smaller statistical errors are obtained. Still, the approach is not yet sufficient to detect post-Newtonian effects in the S2 orbit.

\section{Discussion}
Since more than 10 years two groups have assembled independently astrometric data sets for the stars in the central arcsecond of the Milky Way. The data sets are truly independent. They were obtained at different telescopes with different instruments. The analysis chains used different tools (deconvolution in the NTT/VLT case, PSF-fitting for the Keck data), and the definition of astrometric coordinates is implemented in different ways. While it was reassuring, that \cite{Ghez:2008p945} and \cite{Gillessen:2009p1117} concluded very similarly for $M_\mathrm{MBH}$ and $R_0$, it was not clear that the agreement actually also holds for the underlying data. We were able to show that indeed the most simple assumption - namely that the two data sets only differ in the coordinate system definition - is sufficient to perfectly map the two data sets on top of each other. 

We presented for the first time a combined orbit fit, which however only moderately improves the accuracy by which $M_\mathrm{MBH}$ and $R_0$ can be derived from the S2 data:
\begin{eqnarray}
R_0 &=& 8.34 \pm 0.27 |_\mathrm{stat} \pm 0.5|_\mathrm{sys}\,kpc \nonumber\\
M_\mathrm{MBH} &=& 4.40 \pm 0.27 |_\mathrm{stat} \pm 0.5|_\mathrm{sys} \times 10^6\,M_\odot\,\,.
\end{eqnarray}
For the sake of completeness, we cite also the updated numbers for a fit using not just the combined S2 data, but in addition S1, S8, S12, S13 and S14, the same stars \cite{Gillessen:2009p1117} had used:
\begin{eqnarray}
R_0 &=& 8.28 \pm 0.15 |_\mathrm{stat} \pm 0.29|_\mathrm{sys}\,kpc \nonumber\\
M_\mathrm{MBH} &=& 4.30 \pm 0.20 |_\mathrm{stat} \pm 0.30|_\mathrm{sys} \times 10^6\,M_\odot\,\,.
\end{eqnarray}\\
Besides the expected improvement in $\Delta R_0|_\mathrm{stat}$ from $0.17\,$kpc to $0.15\,$kpc, also the systematic errors have decreased mildly compared to the previous work due to the smaller influence of the S2 2002 data.

We conclude that the combination does not help much in overcoming current limitations. The true value of the two data sets actually is their independence, allowing for cross checks and lending credibility to the results. 

Further substantial improvements in measuring the gravitational potential from Sgr~A* by means of stellar orbits probably will come from improved astrometry on existing data, from longer time lines with the existing instruments and finally from advances in instrumentation.
 
\bibliographystyle{apj}
\bibliography{papers}

\end{document}